# A Predicted Dearth of Majority Hypervolatile Ices in Oort Cloud Comets

C.M. Lisse[1], G.R. Gladstone[2,19], L.A. Young[3], D.P. Cruikshank[4], S.A. Sandford[5], B. Schmitt[6], S.A. Stern[3], H.A. Weaver[1], O. Umurhan[5], Y.J. Pendleton[7], J.T. Keane[8], J.M. Parker[3], R.P. Binzel[9], A.M. Earle[9], M. Horanyi[10], M. El-Maarry[11], A.F. Cheng[1], J.M. Moore[5], W.B. McKinnon[12], W. M. Grundy[13], J.J. Kavelaars[14], I.R. Linscott[15], W. Lyra[16], B.L. Lewis[16,18], D.T. Britt[4], J.R. Spencer[3], C.B. Olkin[3], R.L. McNutt[1], H.A. Elliott[2,19], N. Dello-Russo[1], J.K. Steckloff[20,21], M. Neveu[22,23], and O. Mousis[24]

[1] Space Exploration Sector, Johns Hopkins University Applied Physics Laboratory, 11100 Johns Hopkins Rd, Laurel, MD USA 20723  carey.lisse@jhuapl.edu, hal.weaver@jhuapl.edu, ralph.mcnutt@jhuapl.edu, andy.cheng@jhuapl.edu, neil.dello.russo@jhuapl.edu

[2] Southwest Research Institute, San Antonio, TX, USA 28510  randy.gladstone@swri.org, helliott@swri.org

[3] Southwest Research Institute, Boulder, CO, USA 80302 alan@boulder.swri.edu, layoung@boulder.swri.edu, joel@boulder.swri.edu, colkin@boulder.swri.edu, spencer@boulder.swri.edu

[4] Department of Physics, University of Central Florida, Orlando, FL 32816 dpcruikshank@comcast.net, dbritt@ucf.edu

[5] Astrophysics Branch, Space Sciences Division, NASA/Ames Research Center, Moffett Field, CA, USA 94035 scott.a.sandford@nasa.gov, orkan.m.umurhan@nasa.gov

[6] Université Grenoble Alpes, CNRS, CNES, Institut de Planétologie et Astrophysique de Grenoble, Grenoble, France bernard.schmitt@univ-grenoble-alpes.fr

[7] Space Science and Astrobiology Division, NASA/Ames Research Center, Moffett Field, CA,USA 94035 Pendletonyvonne@gmail.com, jeff.moore.mail@gmail.com

[8] Astrophysics and Space Sciences Section, Jet Propulsion Laboratory/Caltech, Pasadena CA 91109, USA  jkeane@caltech.edu

[9] Dept. of Earth, Atmospheric, and Planetary Sciences, Massachusetts Institute of Technology, Cambridge, MA, USA 02139 rpb@mit.edu, aearle@mit.edu

[10] Laboratory for Atmospheric & Space Physics, University of Colorado, Boulder, CO, USA 80303 mihaly.horanyi@lasp.colorado.edu

[11] Birkbeck, University of London, WC12 7HX, London, UK  m.elmaarry@bbk.ac.uk

[12] Department of Earth and Planetary Sciences and McDonnell Center for Space Sciences, One Brookings Drive, Washington University, St. Louis, MO 63130 mckinnon@wustl.edu

[13] Lowell Observatory, 1400 West Mars Hill Road, Flagstaff, AZ 86001  W.Grundy@lowell.edu

[14] NRC Herzberg Inst of Astrophysics, 5071 W Saanich Road, Victoria BC V9E 2E7 BC, Canada  JJ.Kavelaars@nrc-cnrc.gc.ca

[15] Hansen Experimental Physics Laboratory, Stanford University, Stanford,  CA  94305-9515 linscott@stanford.edu

[16] Dept. of Astronomy, New Mexico State University, PO BOX 30001, MSC 4500, Las Cruces, NM 88003-8001 wlyra@nmsu.edu

[17] Department of Astronomy, Columbia University, 550 W. 120th St., New York, NY 10027 bll2124@columbia.edu

[18] Division of Astronomy and Astrophysics, University of California, Los Angeles, 475 Portola Plaza, Los Angeles, CA 90025

[19] Physics and Astronomy Department, University of Texas at San Antonio, San Antonio, TX 78249, USA

[20] Planetary Science Institute, Tucson, AZ 85719  jordan@psi.edu

[21] Department of Aerospace Engineering and Engineering Mechanics, University of Texas at Austin, Austin, TX, 78712

[22] Department of Astronomy, University of Maryland College Park, College Park, MD 20742  marc.f.neveu@nasa.gov

[23] NASA/Goddard Space Flight Center, Planetary Environments Laboratory, Greenbelt, MD 20771

[24] Aix-Marseille Université, CNRS, CNES, LAM, Marseille, France olivier.mousis@lam.fr

16  Pages, 2 Figures, 1 Table

Proposed Running Title: "**A Predicted Dearth of Hypervolatile Ices in Oort Cloud Comets**"

Please address all future correspondence, reviews, proofs, etc. to:

Dr. Carey M. Lisse

Planetary Exploration Group, Space Exploration Sector

Johns Hopkins University, Applied Physics Laboratory

SES/SRE, Bldg 200, E206

11100 Johns Hopkins Rd

Laurel, MD 20723

240-228-0535 (office) / 240-228-8939 (fax)

Carey.Lisse@jhuapl.edu

**Abstract.** We present new, ice species-specific New Horizons/Alice upper gas coma production limits from the 01 Jan 2019 MU69/Arrokoth flyby of Gladstone *et al.* (2021) and use them to make predictions about the rarity of majority hypervolatile (CO, $N_2$, $CH_4$) ices in KBOs and Oort Cloud comets. These predictions have a number of important implications for the study of the Oort Cloud, including: determination of hypervolatile rich comets as the first objects emplaced into the Oort Cloud; measurement of $CO/N_2/CH_4$ abundance ratios in the proto-planetary disk from hypervolatile rich comets; and population statistical constraints on early (< 20 Myr) planetary aggregation driven versus later (> 50 Myr) planetary migration driven emplacement of objects into the Oort Cloud. They imply that the phenomenon of ultra-distant active comets like C/2017K2 (Jewitt *et al.* 2017, Hui *et al.* 2018) should be rare, and thus not a general characteristic of all comets. They also suggest that interstellar object 2I/Borisov may not have originated in a planetary system that was inordinately CO rich (Bodewits *et al.* 2020), but rather could have been ejected onto an interstellar trajectory very early in its natal system's history.

## 1.   Introduction.

Lisse *et al.* 2021 (hereafter Lisse+21) presented state of the art saturation vapor pressure ($P_{sat}$) and gas production rate ($Q_{gas}$) curves for ices expected in Kuiper Belt Objects (KBOs) like 2014 $MU_{69}$ (hereafter Arrokoth) and Pluto. These curves were informed by the ices found in comets and on the surfaces of outer solar system Centaur, Trojan, and KBO bodies. The primary purpose for calculating these $P_{sat}$ and $Q_{gas}$ curves was to determine which chemical species could be present in abundance in Arrokoth given the 3-sigma non-detection upper limit of H coma gas production of $10^{24}$ mol/sec from New Horizons (NH) Alice instrument airglow observations of Arrokoth (Stern *et al.* 2019). Assuming that the H production upper limit was a good proxy for the production rate limits of ices found in other small solar system icy bodies, Lisse+21 went on to show that there could not be any hypervolatile (e.g., $N_2$, CO, $CH_4$) or mesovolatile ($C_2H_6$, $C_3H_8$, $C_6H_6$, $SO_2$, $H_2S$, etc.) pure ice species in any substantial abundance on the surface of Arrokoth. Rather, refractory hydrogen-bonded ice species such as water, methanol, HCN, and ammonia that remain thermally stable against sublimation into space over Gyrs at the local surface temperatures should be in residence[1]. This finding was consistent with the NH/LEISA and NH/MVIC findings of a surface uniformly rich in tholins, methanol ice and likely water ice (Grundy *et al.* 2020).

Steckloff *et al.* 2021 (hereafter Steckloff+21) used $P_{sat}$ and $Q_{gas}$ curves, coupled with a simple one-dimensional model of comet interior processes (Steckloff *et al.* 2015, 2016), to model Arrokoth's hypervolatile ice content. They found that insolation would deplete Arrokoth of pure hypervolatile ices within the first $10 - 100$ Myr (with little variance due to the amount of short-lived radioactive nuclides), consistent with the lack of hypervolatile activity reported by Lisse+2021. Steckloff+21's results were subsequently confirmed by the ~10 Myr loss time for CO from a 30 km diameter body by Davidsson (2021) and the 24 +/-3 Myr hypervolatile loss time from Arrokoth found by Prialnik (2021).

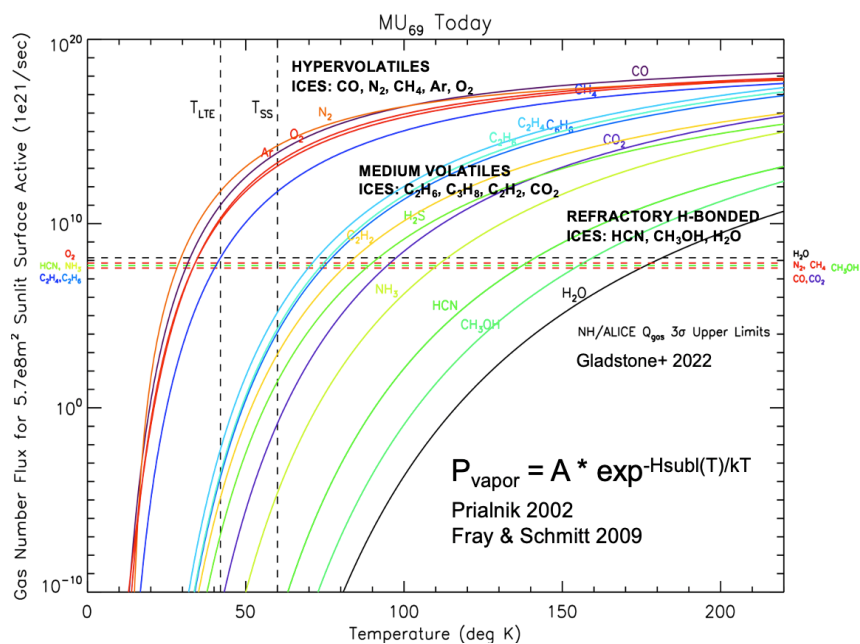

**Figure 1 – Species specific 3-sigma $Q_{gas}$ production upper limits for** Arrokoth, as determined by Gladstone *et al.* 2021 (horizontal colored dashed lines at $10^7 - 10^8$ x $10^{21}$ molecules/sec). Also plotted on the same scale are the gas production rates for different expected icy species found in comets and KBOs (colored curves), as well as the local LTE temperature at 45 au from the Sun for Arrokoth and its sub-solar (noon-time) temperature (vertical dashed lines). Hypervolatile species like $N_2$, CO, and $CH_4$ (red) violate the new NH/Alice upper detection limits by 6-8 orders of magnitude. After Lisse+21.

In a companion paper, Gladstone *et al.* 2022 present new analyses of an NH/Alice appulse observation of the Sun to determine upper limits to gas production rates for the ice species studied in Lisse+21 - and these are all at the ~0.1-1x10²⁹ molecule/sec level, higher than the H-atom proxy number quoted in Lisse+21. These new species specific values, while high compared to the $Q_{gas\_H}$, nevertheless support the arguments made and conclusions drawn in Lisse+21 and Steckloff+21 concerning the lack of pure hypervolatile outgassing activity, and thus lack of pure hypervolatile species present, on Arrokoth in the present day thermal environment. Here we correct and update the literature argument for the non-presence of pure hypervolatile species by presenting the updated individual species limits (Figure 1).

---

[1] To be precise, the curves track the behavior of pure ices and almost pure (> 90% one species) ices. More mixed phases usually have much more complicated behaviors and when such are determined to be present, will require individual study as treatment.

## 2.    A Predicted Dearth of Near-Pure Hypervolatile Ices in the Oort Cloud.

## 2.1    The Origin and Sources of Hypervolatile Ices.

By considering the provenance and very early history of these hypervolatile and refractory ices, we can draw another important solar system ices prediction from these analyses. The hypervolatile ices can be identified as Ehrenfreund (2001)'s small, low-polarity ices, like CO, $N_2$, $O_2$, and $CH_4$. Unable to bind well to dust grain surfaces, such "apolar" ice molecules form mainly from gas-phase reactions and condense directly to ice in cold molecular clouds at extremely cold temperatures (~10-20 K), where thermal energy can be dominated by Van der Waals interactions. By contrast, Ehrenfreund (2001)'s more massive and strongly polar ices, termed here as "refractory" ices, adhere readily to grain surfaces at temperatures upwards of 100K and are thus are efficiently formed as products of heterogeneous grain catalysis on the surfaces of dust grains in clouds.

Often forgotten in solar system studies is that the initial feedstock of gas and dust in the cold molecular cloud (that became the proto-solar nebula), like the bulk Interstellar Medium (ISM), had a dust to gas mass ratio D/G ~ 0.01, 2 orders of magnitude smaller than the D/G ~ 3 $^{+4/-2}$ regime of solid bodies in the solar system today (Ishii *et al.* 2018, Zhukovska *et al.* 2018, Choukroun *et al.* 2020). Similarly, in the very early, gas-dominated solar system, with a solar C/O ratio = 0.54, one would expect ~ 1 $H_2O$ molecule for 1 CO molecule, rather than the 5 to 1000 $H_2O$ molecules for 1 CO we see today in comets (i.e., $Q_{CO}/Q_{H2O}$ = 0.1 to 20%, not ~100%; Bockelee-Morvan *et al.* 2004; Bockelee-Morvan & Biver 2016; Mumma & Charnley 2011). The key point here is that easily-vaporized hypervolatile apolar ice loss in the very early solar system (upwards of 99% of the original gas content) was a normal and expected occurrence concomitant with loss of gas from the circumstellar proto-planetary disk to the ISM at 5 – 10 Myr (Williams & Cieza 2011, Ercolano

& Pascucci 2017), and necessary to produce the small icy outer solar system bodies we know today.

However, the only direct evidence we see for abundant hypervolatile ices in today's solar system is in the atmospheres of the giant planets and their largest moons and on the surfaces of the largest KBOs (Barucci *et al.* 2008, 2011; Brown 2012). These are all bodies capable of supporting gravitationally bound atmospheres (Schaller & Brown 2007, Zahnle & Catling 2017, Young *et al.* 2020). Like Arrokoth, small KBOs and Centaurs show evidence for methanol and water ice, (Cruikshank *et al.* 1998; Barucci *et al.* 2008, 2011) but not hypervolatiles. Comets, which are descended from in-scattered modern KBOs (the short period comets; Lisse *et al.* 2002, Jewitt 2009, Dones *et al.* 2015) or ancient outwardly scattered KBOs (Oort Cloud comets, Brasser & Morbidelli 2013, Morbidelli & Nesvorny 2018; Garrod 2019) do not show any obvious surface absorption features due to ices, except for the rare small patch of water ice or frost (Sunshine *et al.* 2006; Quirico *et al.* 2015, 2016; Fornasier *et al.* 2015, 2019; Lisse *et al.* 2017).

Cometary comae **do** show evidence for low, but finite levels of hypervolatile species (CO at 0.5 – 25% vs water, $N_2$ at 0.1 – 0.3 % vs water, $CH_4$ at 0.2 – 1.0% vs water, [CO + $CO_2$] at ~20% vs water; Bockelee-Morvan *et al.* 2004; A'Hearn *et al.* 2012; Bockelee-Morvan & Biver 2016; Mumma & Charnley 2011). This implies that there must be some reservoir for them in these objects – but it cannot be in deeply held near-pure hypervolatile ice phases. Not only is this highly unlikely given the study of cometary thermal timescales for ~1-10 km sized cometary nuclei with lag deposits published by Davidsson 2021 and Steckloff+21, bolstered by the modeling of Arrokoth's ice loss performed by Prialnik (2021) – it would take deeply buried regions held at ~15 K since the beginning of the Solar System 4.56 Gyr ago, coupled with an enormous, aphysical lag delay, to force this to happen, contrary to the modern 30 - 40K core temperatures of Arrokoth determined by Davidsson 2021, Lisse+2021, Prialnik 2021, Steckloff+2021, and Umurhan *et al.* 2022[2]. The existence of deeply buried pure hypervolatile ices is also contrary to the lack of substantial hypervolatile emission, such as CO, from end member objects 45P/Honda-Mrkos-

---

[2] Another thermal loss timescale on the order of 4 Gyr has also recently been published by Kral *et al.* 2021, but this model utilized an aphysical thermal diffusivity = $10^{-10}$ m²/s which is more than 2 orders of magnitude lower than that of the most insulating solid material currently known, aerogel, and more than 3 orders of magnitude lower than the parameter values used by Davidsson21, Prialnik21, and Steckloff21 in their models. Scaling the 4 Gyr timescale by a factor of 1/300 to correct for the thermal diffusivity error, the Kral21 timescales become consistent with the 10-20 Myr timescales for loss of all pure CO/$N_2$/$CH_4$ -like ices.

Pajdušáková, 46P/Wirtanen, and 103P/Hartley 2 (small comets near the end of their lives emitting chunks of their cores, A'Hearn *et al.* 2011, DiSanti *et al.* 2017. Steckloff & Samarasinha 2018). [E.g., a massive landslide is thought to have exposed the interior of Comet 103P/Hartley 2 only ~3 decades ago (Steckloff *et al.* 2016), however the EPOXI mission detected no evidence of strong hypervolatile emission (A'Hearn *et al.* 2011)].

Neither was any marked increase in hypervolatile emission seen from the recently split comets 73P/SW3 (Dello Russo *et al.* 2007) nor from 17P/Holmes (Dello Russo *et al.* 2008), again showing from direct observational studies that the cores/interiors of Jupiter Family Comets do not contain any large amount of pure hypervolatile ices "hidden down deep"[3]. Instead the low but finite level of hypervolatile ices seen in cometary comae is thought to be due to their inclusion as minority impurities in bulk majority water ice and $CO_2$ ice phases (Jewitt 2009, Lisse+21, Davidsson 2021), and it is from these phases that it can be released via thermal, sputtering, or collisional processes (Kral *et al.* 2021 and references therein).

## 2.2    Small Outer Solar System Body Hypervolatile Ice Evolution.

The lack of abundant near-pure hypervolatile ices in outer solar system bodies, except the very largest with gravitationally bound atmospheres, can be explained by thermal processing effects. After the initial condensation of ices and small icy bodies in the early molecular cloud/T Tauri star/Protoplanetary disk phases, another important epoch of the early solar system was the so-called "disk clearing" time, when enough of the gas (~90% or  more) was removed from the solar system's T-Tauri accretion disk for it to become optically thin to optical radiation out as far as the Kuiper Belt. In Lisse+21 this was termed "morning coming to the Kuiper Belt", and by Steckloff+2021 as the dawn of the "Sublimative Period" of the Kuiper Belt. Lasting for the first

---

[3] Instead, we follow the arguments of Iro *et al.* (2003) and Jewitt (2009) that the hypervolatiles and moderately-volatile species in Centaurs and SP comets are protectively stored in $H_2O$ ice matrices – first at high concentrations in cold ($T < 80K$) amorphous water ice composites, then in lower concentrations in warmer $T > 100K$ crystalline water ice matrices limited by the maximum interlattice "pore space" trapping capability of the crystallite. Li *et al.* 2020, using HST, verified the lack of activity of Centaurs beyond 10 AU. This is consistent with the early work of Prialnik *et al.* (1987) who argued that the presence of amorphous ice within the subsurface of comets – inferred from observations of outgassing at surprisingly large heliocentric distances (5.8 - 11.4 AU) and attributed to the annealing of amorphous ice as comets first enter the inner Solar System (Prialnik & Bar-Nun, 1990; 1992; Meech *et al.* 2009) – provides a clear constraint on the maximum parent body temperatures ($T < 135$ K) experienced over long ($\gg 100$ Myr) comet lifetimes. The total Q(CO + CO2) of comets is also suspiciously close to the ~20 % carrying capacity of crystalline water ice – A'Hearn *et al.* 2011.

10-50 Myr of the solar system (~20 My for Arrokoth but depending on object size; Steckloff+2021, Prialnik2021), the Sublimative Period led to a sudden spike in the local surface insolation temperatures near to what they are today, and a new wave of hypervolatile ice sublimation as direct insolation became the dominant steady state surface heat source.

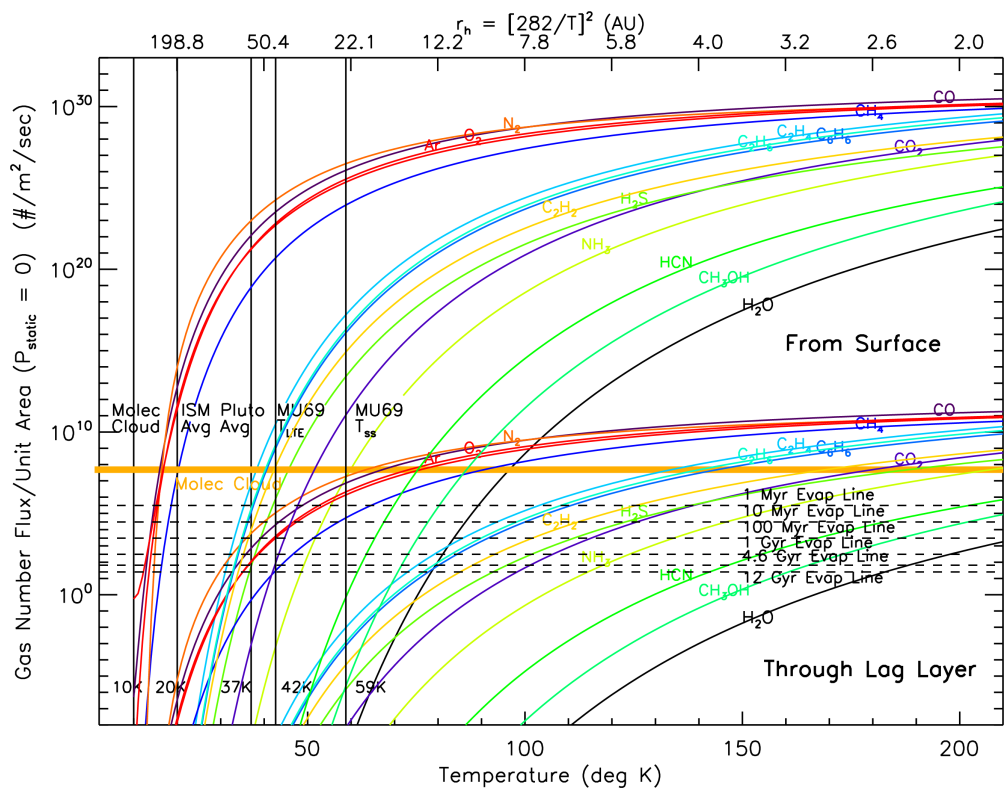

**Figure 2 – Species specific Q$_{gas}$ vs Temperature curves for species expected in comets and KBOs (upper colored curves).** Horizontal dashed lines: values of the thermally driven outgassing rates at which an icy species is depleted in 1,10,100, 1000, 4600, and 12,000 Myrs for an Arrokoth-sized body. **Upper colored curves**: loss rates of a piece of surface ice evaporating into free space. **Lower colored curves**: the much slower loss rates for the same species, after allowing for an overlying lag layer with thermal diffusivity = 3x10$^{-7}$ sec$^2$/m impeding the flow of heat and gas into free space from the interior (Davidsson21, Steckloff 21, Prialnik21). Top axis: heliocentric distance from the Sun for a blackbody at local thermal equilibrium temperature T. From these curves and constraints, one can see that hypervolatile ices CO, N$_2$, and CH$_4$ are stable in cold, dense molecular clouds and in modern KBOs residing beyond ~100 AU from the Sun, but were lost by ~20 Myr after Arrokoth's formation. One can also see that meso-stable ices like CO$_2$ , while easily removed from the surface, can remain stably at depth inside a KBO for more than the age of the solar system.

This is where the curves in Fig. 2 come in. Using the same Q$_{gas}$ curves and the insolation energy balance relation T$_{LTE}$ = [(1-A)/ε]$^{1/4}$ 282/sqrt(r$_h$) $^{o}$K for a non- or slowly-sublimating body with average albedo A and emissivity ε across the solar spectrum, Lisse+21 showed the different spatial regions of stability for the different expected ices (Table 1). As the temperatures of these bodies are raised, the first species to sublime are the most volatile (i.e., the ones that evaporate at the lowest temperatures), and if present at high enough relative abundances will sublimatively cool the nucleus and keep its surface temperatures "pinned" near the temperature at which the ice species flash from solid to vapor (i.e., the near-vertical portions of the $Q_{gas}$ curves shown in Fig. 2, where the vapor pressure of a given species increases by 20 orders of magnitude in 10 deg K).

Given this, and allowing for slower loss rate indicated by Davidsson21, Prialnik (2021), and Steckloff+21 for sublimation from the deep interior of a realistic Arrokoth-like body with inactive lag deposits overlaying the actively subliming ices, one sees that the hypervolatile ices are only stable on Gyr geological and astronomical timescales at heliocentric distances of 100 AU and beyond, in the outer extremes of the Kuiper Belt and in the Oort Cloud region (Fig. 2)[4]. There is simply no way to insulate these cryogenic species versus loss over 4.56 Gyr, given materials known to humankind for KB objects with dayside/nightside surface temperatures of 60/30 K (Bird *et al.* 2022) and interior core temperatures on the order of 40K (Lisse+2021, Umurhan *et al.* 2022).

**Table 1 - Ice Stability vs Sublimation Regions in the Solar System**

| Solar System Region | $T_{LTE}$ Range[a] (degrees K) | Crystalline Water Ice | Amorphous Water Ice | Other Refractory Ices ($CH_3OH$, HCN, $NH_3$,$H_2CO$, etc.) | Hypervolatiles ($N_2$, CO, $CH_4$, etc.) |
|---|---|---|---|---|---|
| Inner System ($r_h$ < 2.5 AU) | 180 - 5780 | Not stable | Not stable | Not stable | Not stable |
| Outer Asteroid Belt to Saturn ( 2.5 < $r_h$ < 10 AU) | 90 - 180 | Stable | Not stable | Not stable | Not stable |
| Saturn to Uranus (10 < $r_h$ < 20 AU) | 60 - 90 | Stable | Stable | Not stable | Not stable |
| Uranus thru Kuiper Belt (20 < $r_h$ < 70 AU) | 35 - 60 | Stable | Stable | Stable | Not stable |
| Outer Kuiper Belt, Oort Cloud ($r_h \geq$ 100 AU) | 10-20[b] | Stable | Stable | Stable | Stable |

[a] - For a spherical, isothermal greybody. A large, finite, rotating body with different optical and infrared emissivities will have localized surface regions, like the equator and sub-solar (noon) point, that will be warmer than $T_{LTE}$. In other words, surface temperatures will vary higher and lower than $T_{LTE}$ depending on the real physical and rotational structure of the body, pushing the instability regions outward from the simple isothermal greybody case.

[b] - For the Oort Cloud, the Sun's energy input due to insolation is unimportant compared to energy input from the local galactic ISM heat bath of 10 - 20 K. Oort Cloud comets spend > 99.9% of their time in the galactic ISM.

---

[4] This simple long-term thermal analytical line of reasoning is valid because other processes, like the transient heating introduced via galactic cosmic rays, nearby passage of O/B stars, explosion of nearby supernovae, and perihelia passages (Stern 2003), are strictly "surface processes" that only affect the top few meters of an object (Lisse+21).

## 2.3    Late Oort Cloud Formation Implies a Lack of Bulk Hypervolatile Ices.

However, compared to observations this presents a puzzle, given that the only Oort Cloud comets known to emit majority hypervolatiles (but little to no refractory volatiles) are Comet C/2016 R2 (Biver *et al.* 2018, McKay *et al.* 2019), and perhaps the new hyper-distant active comets such as C/2017 K2 PANSTARRS (Jewitt *et al.* 2021, Yang *et al.* 2021). This handful of comets represents a negligible ($\sim 10^{-3}$) fraction of all the known Oort Cloud comets.

The answer lies in the formation timescale for the Oort Cloud. Unlike the Kuiper Belt, which is at or near to the edge of the original PPD, the Oort Cloud is a later construct, formed of billions of icy planetesimals that failed to aggregate onto and become part of one of the giant planets - but were instead scattered out onto nearly hyperbolic, barely bound, million year orbits (Boe *et al.* 2019). Originally thought to be created quickly during the 1-10 Myr time of greatest mass growth of the giant planets (Duncan *et al.* 1987), later work has shown (Stern 2003; Dones *et al.* 2004, 2015; Brasser 2007), that it is very difficult to launch a small planetesimal onto a Myr long, near-unity eccentricity orbit through a midplane dense with gas and other planetesimals, rather than scattering them off of other bodies nearby in the ecliptic. Most current models (e.g., Brasser & Morbidelli 2013, Morbidelli & Nesvorny 2018; Garrod 2019) favor instead population of the Oort Cloud to start around the time of the giant planet orbital instability, at 100's of Myrs after the beginning of the solar system, via Neptune's scattering of planetesimals from the Kuiper Belt into the Cloud. The same dynamical processes also removed upwards of 99% of the mass of the Kuiper Belt through ejection and accretion, and created the "dynamically hot" scattered disk KBOs seen in the modern Kuiper Belt (see Nesvorný 2018).

As mentioned above and shown in Fig. 2, any small icy solar system body found in regions from the Kuiper Belt inward, including the giant planet region, will lose, via isolation heating, its hypervolatile majority ices to vacuum within 10's of Myrs. Since current models have > 99% of icy Oort Cloud objects emplaced after 100's of Myr's worth of residence time in the giant planet or Kuiper Belt regions, they will have lost their bulk majority hypervolatiles, so we can expect that the large majority of Oort Cloud objects will lack bulk hypervolatile species like CO, $N_2$, and $CH_4$.

## 2.4    An Illustrative Counterexample?  Comet 2016/R2.

There could be exceptions to the rule of Oort Cloud hypervolatile-depleted comets. For example, a few (< 1%) very extraordinary bodies  may have been quickly (in < 20 Myrs' time) inserted from the giant planet region. These few bodies, if large ($R_{nuc}$ > 5 km), can then endure thousands of orbits' worth (i.e., Gyrs) of hypervolatiles loss upon perihelion passage. (But again, the vast majority, >99% of the icy planetesimals launched into Oort Cloud orbits from the giant planet region, after a ~100 Myr delay, should contain no icy volatiles, except as minority impurities in refractory ice matrices, reproducing the apparent situation we observe today).

As suggested in Lisse+21, comet C/2016 R2 may represent just such an important example of an object outgassing as one would expect if it was rich in nearly-pure $N_2$, CO, and $CH_4$ ice. Such an object would show large amounts of these species in its coma as it moved from the Oort Cloud to inside the Kuiper Belt (< 30 AU), but would show *no trace* of the most refractory ice species - water - in the coma vapor phase, since the ~ 20K temperature of the comet's surface enforced by the hypervolatiles' latent heat of sublimation means that the vapor pressures of refractory ices, as with the usually dominant water, are negligible - and will remain so until the available hypervolatile solids are exhausted. The lack of water gas production from R2 is very telling  – for such an active comet, with $Q_{gas}$ ~ $Q_{CO}$ = $10^{29}$ mol/sec[5], $H_2O$ is ***always*** detected in a comet. Adopting the 1.1 x $10^{28}$ mol/sec water upper limit of Biver *et al.* 2018, and using the observed $Q_{CO}$ production rate of ~1.1 x $10^{29}$ mol/sec and the $Q_{N_2}/Q_{CO}$ ratio of ~8%, we see that $Q_{CO}/Q_{H_2O}$ > 10. The $Q_{CO}/Q_{H_2O}$ is at least 5 times higher than in any other comet observed inside the water ice line at ~2.5 AU, including disrupted/fragmented/hypervolatile comets exposing their core regions (Section 2.1). Similarly, $Q_{N_2}/Q_{H_2O}$ > 0.8 is at least 16 times higher than in other comets, making the simple "comet with strange compositional abundance argument" suggested by Biver *et al.* (2018) hard to accept for this object. Instead, the observed $Q_{N2}/Q_{CO}$ ratio, indicative of the relative coma abundance of $N_2$ vs CO, is close to the solar N: C atomic abundance ratio (Anders & Grevesse 1989, Lodders 2003, Asplund *et al.* 2005) as one would expect for a mix of PPD ices of true solar abundance.

---

[5] Note that a $10^{29}$ molecules/sec level of hypervolatile outgassing is reasonable, and can be supported by a 15 km radius object that is losing molar (~$10^{15}$ molecules/cm²) amounts of ice surface every second : $(2\pi R_{nuc}^2)*(2x10^{15}/cm^2/sec) \sim 3x10^{28}$ molecule/sec. From Lisse+21, the evaporation of 1 mole of CO or $N_2$ ice requires ~7.3 kJ of heat, thus the amount of hypervolatile cooling from the emission rate of $10^{29}$ molecules/sec = 1.6 x $10^5$ moles/sec is ~ 1.4 x $10^9$ J/sec, the same order of magnitude, (1-0.9) * $\pi R_{nucleus}^2$ * (0.1W/cm² * (1.0 AU/2.6 AU)²) = 1 x $10^{10}$ W, as the insolation heating of the Sun is delivering to R2's nucleus (with assumed albedo = 0.90) at 2.6 AU. So the observed mass loss rate of R2 being attributed to hypervolatile sublimation makes rough sense if R2 is feverishly sublimating from its entire sunlit surface during the small portion of each 20,000 yr long orbital cycle where it is intensely heated. For a 15 km radius object of ~0.5g/cm³, with ~7 x $10^{15}$ kg total mass and current orbital loss rate of ~ 3 x $10^7$sec [1 yr]*(28 amu for CO/$N_2$*1.67 x $10^{-27}$ kg/amu)*1 x $10^{29}$ molecule/sec ~ 1 x $10^{11}$ kg should be able to endure ~7 x $10^4$ more of these kind of passages before dissipating, comprising another or ~2 x $10^4$ yrs/passage * 7 x $10^4$ passages = 1.4 Gyrs.

## 2.5    A Possible Additional Small Population of Large-KBO Derived Comets.

Another class of hypervolatile rich Oort Cloud insertion models, due mainly to Desch & Jackson (2021), suggests that free-flying hypervolatile rich fragments of dwarf planet surfaces (such as Pluto's) can be created via energetic impacts during the 2:1 Jupiter:Saturn resonance epoch, and then scattered into the Oort Cloud along with multitudes of "normal" KBOs. Again the hypervolatile rich surface fragments will be a very small minority compared to the bulk hypervolatile stripped implanted KBO objects (~0.1%), and the ratio $N_{Oort\ hypervolatile\ rich}/N_{Oort\ normal\ comet}$ for this process would be determined by the efficiency of carving out km-sized chunks of differentiated dwarf planet surface and ejecting them into the Kuiper Belt. This ratio could thus be very different than that determined by the low efficiency of dynamically scattering very young (< 20 Myr) planetesimals by the giant  planets through the busy, well-populated disk of the very young solar system.

## 3.    Conclusions and Observational Tests.

The prediction that Oort Cloud comets should be depleted in majority species hypervolatiles has some important implications for the study of the Oort Cloud. Many of these are directly testable, and include:

**(a)** - The prediction that hypervolatile rich comets are rare, and thus that abundant hypervolatiles will not be a general characteristic of all comets, despite the current high levels of interest expressed over the phenomenon of ultra-distant active comets like C/2017K2, (Jewitt *et al.* 2017, Hui *et al.* 2018).

**(b)** - If the hypervolatile rich objects came from the lucky few primitive planetesimals scattered by the giant planets, then they must have been emplaced within ~20 Myr (Davidsson21, Prialnik21, Steckloff+21), and thus represent the first objects in the Oort Cloud, and can potentially provide a direct measurement of $CO/N_2/CH_4$ ratios in the proto-planetary disk.

**(c)** - Obtaining good, debiased statistics on the frequency of hypervolatile rich Oort Cloud comets (i.e., from $N_{Oort, \text{hypervolatile rich}}/N_{Oort, \text{normal comet}}$) can provide important constraints on models of early (< 20 Myr) versus later (0.05-2.0 Gyr) emplacement of objects into the Oort Cloud from the giant planet region of the solar system.

**(d)** - If instead, following Desch & Jackson (2021), the hypervolatile rich Oort Cloud comets are sourced from differentiated KBO surfaces, then they could be roughly coeval with the rest of the Oort Cloud and their frequency $N_{Oort, \text{hypervolatile rich}}/N_{Oort, \text{normal comet}}$ will reflect the proportions of KB dwarf planet surfaces in the era of giant planet instability containing hypervolatile rich phases, convolved with the efficiency of launching these phases into the Oort Cloud.

**(e)** - The presented arguments also provide an alternate explanation for the high observed CO levels in interstellar object 2I/Borisov than the one presented in Bodewits *et al.* 2020 – i.e., that rather than coming from a hypervolatile CO rich system versus ours, 2I could instead simply have been launched early (within the first 20 Myr or so) of its natal system's history, or is representative of the natal system's dwarf planet surfaces.

## 4.    Acknowledgements.

The authors would like to thank NASA for financial support of the New Horizons project that funded this study, and the entire New Horizons mission team for making the success of the flyby and its fantastic data return possible. The authors are also indebted to an anonymous graduate student who attended the "UCLA Planet Lunch Seminar" on 07 May 2021 for the inspiration to write this material up into an article.

# 5. References.